\documentstyle[12pt,aaspp4]{article}

\slugcomment{Accepted for publication in the Astrophysical Journal}

\lefthead{Sankrit \& Wood} \righthead{Resonance Line Scattering in Shocks}

\begin{document}

\title{Resonance Line Scattering in Supernova Remnant Shocks}

\author{Ravi Sankrit\altaffilmark{1}}
\and
\author{Kenneth Wood\altaffilmark{2}}

\altaffiltext{1}{Department of Physics and Astronomy,
The Johns Hopkins University,
3400 N. Charles St., Baltimore, MD 21218}
\altaffiltext{2}{Harvard-Smithsonian Center for Astrophysics,
60 Garden St., Cambridge, MA 02138}


\begin{abstract}

We present a three dimensional radiative transfer model to examine the
effects of resonance line scattering in the post-shock flow behind a
non-radiative supernova remnant shock.  For a rippled shock front
viewed edge-on, line scattering significantly reduces the observed flux
of C~IV~$\lambda$1549 and N~V~$\lambda$1240, two important diagnostic
lines in the ultraviolet spectra of supernova remnants.  The correction
factor (defined to be the ratio of the line flux that would be observed
neglecting scattering, to the actual observed line flux) is a function
of position within the filament.  For sufficiently large regions that
include crisp edges as well as more diffuse regions of the filament
structure, the C~IV and N~V correction factors are between about 1.5
and 3.5 (and the C~IV correction factor is invariably larger than the
N~V correction factor).  The correction factors have a larger range
when smaller regions are considered.  The C~IV correction factor is
about 6 at the filament edges, while the N~V correction factor is about
4.  These simulations of resonance line scattering will be useful for
the analysis of supernova remnant shock spectra.

\end{abstract}



\section{Introduction}

The optical and ultraviolet emission from middle-aged supernova
remnants (SNRs) arises mainly in shock excited gas.  The shocks are
driven by the supernova blastwave into surrounding interstellar
clouds.  Several lines from a range of ionization states are seen in
the spectra of these shocks.  The line strengths can be used to obtain
properties such as the shock velocity, the pre-shock density and the
elemental abundances (e.g.,~\cite{ray91}).

A standard method of interpreting SNR spectra is to compare observed
line fluxes with shock model predictions, and thereby derive the shock
properties.  \cite{ray80} pointed out that radiative transfer in
resonance lines (permitted lines to the ground state) was a potential
problem in comparing observations with model calculations.  Their
models showed that the optical depths of resonance lines are of order
unity in the flow direction and much higher along the line of sight
when the shock is viewed edge-on.  They found that models which
correctly predicted the observed intensities of intercombination lines
in an \textit{International Ultraviolet Explorer} (IUE) spectrum of a
filament in the Cygnus Loop SNR severely underpredicted the observed
resonance line intensities.  They ascribed this discrepancy to
resonance line scattering in the filament.  The same discrepancy
between observed and predicted strengths of resonance lines, relative
to non-resonance lines was found in other studies of SNR shocks as well
(\cite{ray81}, \cite{ray88}, \cite{ray97}).  Resonance line scattering
effects were directly observed by \cite{cor92}.  In images of a large
field in the Cygnus Loop taken with the \textit{Ultraviolet Imaging
Telescope} (UIT), they found that the C~IV~$\lambda$1549 emission was
significantly reduced near sharp optical filaments, which are the
locations where the shock is viewed close to edge-on (\cite{hes87}).

Thus, it has been known for some time that the observed fluxes of
resonance lines have to be corrected for optical depth effects when
analyzing SNR shock spectra.  The approach towards this problem has
been, for the most part, to obtain shock velocities and pre-shock
densities based on non-resonance lines and then to calculate the
effects of resonance line scattering by comparing the observed fluxes
with the fluxes predicted by a ``best-fit'' model.  That is, the
resonance line strengths themselves were not used in the analysis.  In
one study of filaments in Vela and in the Cygnus Loop, \cite{ray81}
calculated line optical depths from column densities predicted by shock
models.  Having assumed a thin sheet geometry viewed edge-on, and
single scattering, they found attenuation factors for several resonance
lines.  These values, as they point out, are extreme because of the
assumed geometry.  \cite{ray81} further state that quantitative
interpretation of the resonance line strengths is extremely difficult
because they depend on details of the geometry as well as the shock
properties.

Resonance line scattering also affects the ratio between the line
strengths of the components of a resonance doublet such as
O~VI~$\lambda\lambda$1032,1038.  If the emission is optically thin,
then the observed ratio, $I_{1038}$/$I_{1032}$ should be 0.5.  For
higher optical depth, the ratio will increase and in the optically
thick case, will be equal to 1.0.  The O~VI doublet was partly resolved
in a \textit{Hopkins Ultraviolet Telescope} (HUT) spectrum of a
non-radiative filament in the Cygnus Loop (\cite{lon92}).  The observed
ratio $I_{1038}$/$I_{1032}$ was 0.6, indicating an optical depth of
about 1 for the 1032~\AA\ line, and a reduction of the total O~VI flux
by a factor of 1.4.  The use of this method depends on having
sufficient spectral resolution to resolve the doublet components.
Furthermore, it has long been known from numerical models that the
emission from SNR shocks is highly stratified.  The O~VI emission does
not trace lower ionization species such as N~V and C~IV.  Therefore the
O~VI~$\lambda\lambda$1032,1038 optical depth cannot be used to derive
the optical depths of lines such as N~V~$\lambda$1240 and
C~IV~$\lambda$1549.

In this paper, we introduce a method using a Monte Carlo transfer code
to calculate the effect of resonance line scattering in SNR shocks.
The effort has been motivated in part by the spatially resolved
ultraviolet spectra of SNRs that can be obtained by new instruments
such as the \textit{Space Telescope Imaging Spectrograph} (STIS) on
board the \textit{Hubble Space Telescope} (HST) and the \textit{Far
Ultraviolet Spectroscopic Explorer} (FUSE).  An example of such spectra
are those of a non-radiative shock in the Cygnus Loop obtained with
STIS.  (This is the same filament observed by \cite{lon92}, referred to
above.) One spectrum, taken with the spatial axis of the spectrograph
slit placed perpendicular to the shock front, shows the ionization
stratification of the post-shock gas (\cite{san00}).  The other spectra
have been taken with the slit placed parallel to the shock front at
different positions, providing a map of the C~IV and N~V emission in
the post-shock flow.  These two lines are the only strong lines seen in
the STIS spectra.  In order to interpret the data, it is necessary to
treat resonance line scattering more accurately than has so far been
done.

We base the shock model on a well studied non-radiative shock in the
north east region of the Cygnus Loop.  The shock is assumed to have a
rippled, thin sheet geometry as seen in a WFPC2 H$\alpha$ image
(\cite{bla99}).  The properties of the gas in the post shock flow are
specified using shock model predictions (SBRL).  The shock model is
calculated in one dimension and applied at every point on the rippled
surface to generate the three dimensional density distribution.  In \S2
we describe the Monte Carlo code and then details of the model
filament.  The results from the scattering calculations are presented
in \S3. Finally, In \S4 we discuss the implications of our work, and we
suggest ways in which the method can be used in future studies of SNR
shocks.


\section{Model Description}

The Monte Carlo code we use for our simulations is based on the one
described by \cite{cod95} and has been modified to run on a three
dimensional Cartesian grid (\cite{woo99}).  The code tracks energy
packets as they are scattered within a density grid.  For a total
luminosity $L$ of the grid, N energy packets are tracked over a time
interval $\Delta t$; so each packet has an energy, E$_{\gamma}$~=~$L
\Delta t $/N\@.  Hereafter, we refer to an energy packet as a
``photon''.  The emissivity of a given line and the number density of
the species that emits the line are provided as functions of position.
The code calculates the surface brightness as seen from a specified
viewing direction.  The code does not treat radiative transfer within
the line, instead it considers all photons to be emitted and resonantly
scattered at a single wavelength.  This approximation is valid for
moderate optical depths where photons are not scattered out of the
doppler core of the line.  The code includes forced first scattering
(\cite{wit77}) and a ``peeling off'' procedure (\cite{yus84}) which
allows us to construct model images of three dimensional structures
from any specified viewing angle very efficiently.

We use the non-radiative shock in the northeast limb of the Cygnus Loop
studied most recently by \cite{bla99} and SBRL as the physical basis
for our model.  The H$\alpha$ emission from a non-radiative shock is
due to collisional excitation of neutrals in a narrow zone ($\sim
2\times 10^{14}$~cm) behind the shock front.  Therefore, the WFPC2
H$\alpha$ image presented by \cite{bla99} shows us the morphology of
the shock front.  The essential morphological features of the shock
front can be captured by modeling the rippled sheet geometry with a
simple superposition of sine waves.  We obtained the equation for the
shock front geometry by trial and error, visually comparing with the
WFPC2 H$\alpha$ image.  With lengths in code units ($R_0 =
10^{15}$~cm), the equation used is:
\begin{equation}
z = 3.5\times(2\sin\frac{2\pi(x-200)}{1000} + \sin\frac{2\pi x}{500})
       \times(2\sin\frac{2\pi y}{1000} + \sin\frac{2\pi y}{500})
\end{equation}
Thus, the shock front lies approximately in the $xy$ plane.  The
physical dimensions of the grid used for the calculations are
$6\times10^{17}$~cm along each of the $x$ and $y$ axes and
$6\times10^{16}$~cm along the $z$ axis (the direction of the shock
flow).  The grid has $301\times 301\times 301$ points, so the structure
of the post-shock gas ($z$ direction) is resolved at a scale of
$2\times 10^{14}$~cm.  (At a distance of 440~pc to the Cygnus Loop,
this corresponds to 0\farcs03.)

The lines we are primarily interested in are C~IV~$\lambda$1549 and
N~V~$\lambda$1240, both of which are strong resonance lines detected in
the STIS spectrum presented by SBRL\@.  We specify the emissivities of
these lines and the number density of C$^{3+}$ and N$^{4+}$ based on
the fiducial model for the non-radiative shock presented by SBRL\@.  The
shock has a velocity of 180~km~s$^{-1}$ and the pre-shock hydrogen number
density is 2~cm$^{-3}$.  The carbon and nitrogen abundances are 8.40
and 7.90 on a logarithmic scale where the Hydrogen abundance is 12.00.
The shock model is truncated at a swept up hydrogen column of
$10^{17}$~cm$^{-2}$.  The shock model predictions are combined with the
sine wave sheet to produce three dimensional maps of the line
emissivities and number densities of the relevant ionic species.  The
opacities are derived from average cross sections for thermally
broadened profiles.  The post-shock gas temperature for a
180~km~s$^{-1}$ shock is about $4.2\times10^{5}$~K, assuming that
electrons, protons and ions are in thermal equilibrium.  Using this
temperature, and oscillator strengths of 0.191 and 0.156 for
C~IV~$\lambda$1549 and N~V~$\lambda$1240, respectively (\cite{mor91}),
the cross sections are $\sigma_{C~IV}\times R_0 = 32.50$~cm$^{3}$ and
$\sigma_{N~V}\times R_0 = 22.94$~cm$^{3}$ (note: $R_0 = 10^{15}$~cm).

The approximation made in the simulations that photons are scattered at
a single frequency is valid for moderate line optical depths.  For N~V
and C~IV, scattering outside the doppler core becomes important at
optical depths of about 10,000.  We now estimate the line optical
depths, $\tau_{i}~\sim~\sigma_{i}n_{i}l$, where $\sigma_{i}$ is the
cross-section (given above), $n_{i}$ is the ion density and $l$ is the
path length through the emitting region.  For the filament viewed edge
on, $l~\sim6\times 10^{17}$~cm.  The shock models predict that the
maximum density of C$^{3+}$ is 0.0017~cm$^{-3}$ and of N$^{5+}$ is
0.0006~cm$^{-3}$ in the post-shock flow.  For these maximum densities,
$\tau_{C~IV}~\sim~30$ and $\tau_{N~V}~\sim~13$.  Note that the typical
optical depths through the filament will be smaller than these values,
since the average ion densities are much lower.  Scattering out of the
doppler core of the lines is not important at these optical depths.
Furthermore, the optical depths perpendicular to the sheet are much
smaller.  We find from the shock model that $N_{C~IV} = 2.7\times
10^{12}$~cm$^{-2}$ and $N_{N~V} = 3.6\times 10^{12}$~cm$^{-2}$
perpendicular to the shock front.  These column densities correspond to
optical depths $\lesssim 0.1$ for both lines.  Photons can escape quite
easily from the sheet in the perpendicular direction, making scattering
in the wings even less of an issue.  Velocity gradients along the line
of sight through the emitting gas can also lead to scattering in
frequency.  In our model filament, the shock is moving perpendicular to
the line of sight.  The velocity spread due to the rippled geometry is
of order the thermal velocity width --- there are no velocity gradients
that would make escape of line photons in the wings significant.  In
summary, we have constructed a model filament where the assumption of
scattering at a single wavelength is valid.

The model neglects extinction by dust.  To see if this is justified, we
estimate the dust opacity along the line of sight through the
filament.  We assume a standard ISM dust to gas ratio in the filament
because the non-radiative shock has not yet had time to destroy
grains.  Equation 7-23 of \cite{spi78} gives the relationship between
the projected area of dust grains per H atom, whence we obtain
$\tau_{dust}~=~10^{-21}\times n_{\rm{H}}\times l\times Q_{e}$.  The
post shock hydrogen density, $n_{\rm{H}}~\sim~10$~cm$^{-3}$ is
predicted by the shock model.  The path length through the filament,
$l~=~6\times 10^{17}$~cm, as above.  $Q_{e}$ is the extinction
efficiency factor which is about 2 in the far-ultraviolet (Figure 7.1
of \cite{spi78}).  For these values we obtain $\tau_{dust}~\sim~0.01$
for a line of sight through the filament.  Even if a line photon were
to traverse the length of the filament several times, the dust opacity
is so low that extinction by grains within the filament can indeed be
neglected.


\section{Results}

In Figure~\ref{figmodim} we present model images of the shock as seen
in H$\alpha$, C~IV~$\lambda$1549 and N~V~$\lambda$1240 emission.  Each
model has been constructed with $10^7$ photons emitted towards the
observer (see Appendix in \cite{woo99} for details of the photon
weighting scheme).  The models are calculated for a viewing angle of
$\theta$=90\arcdeg, $\phi$=20\arcdeg\ in standard polar coordinates
(i.e.~tangential to the shock front and 20\arcdeg\ from the $x$-axis).
In these images the value at each point is the number of photons
detected along that line of sight.  The H$\alpha$ model assumes uniform
emission over a very narrow zone ($2\times 10^{14}$~cm) behind the
shock front and, since it is not a resonance line, includes no
scattering.  The H$\alpha$ model image (Figure~\ref{figmodim}a) shows
the morphology of the shock front.  The C~IV~$\lambda$1549 image shown
in Figure~\ref{figmodim}b is what would be observed if there was no
scattering.  Figure~\ref{figmodim}c is the C~IV image with scattering
taken into account.  Figures~\ref{figmodim}d and \ref{figmodim}e show
the same for N~V~$\lambda$1240.  We use the terminology ``thin'' and
``thick'' for the two images shown for each line.  The thin image is
produced by running the code, while assuming no scattering.  This gives
a map of the intrinsic emission from the filament.  The thick image
includes scattering and is a map of the observed emission.  (Note that
the term thick is used for convenience and is not a statement about the
optical depth of the filament.  The line optical depth along any given
line of sight through the filament can be low even in the thick image.)
The results that we present in this section apply only to the model
filament described in \S2, and only for the specific input photon
number and viewing geometry given above.

Each pair of thin and thick images have been scaled logarithmically
between the same minimum and maximum values so they can be visually
compared.  (All the whitespace outside of the filament have values
identically equal to zero.)  An inspection of the thick and thin images
for each line shows that scattering decreases the intensity
significantly along the bright edges of the shock structure.  The
images also show that the N~V emission is more extended than the C~IV
emission all along the front as expected from shock models
(e.g.~Figure~5 of SBRL).  Note that because the N~V emission is so
extended, a small segment (at the trailing edge on the right) has been
truncated in the models.  Less than 2\% of the total photons are lost
because of the truncation and our results are not compromised.

For the selected viewing angle, the model filament is $6\times
10^{17}$~cm long (parallel to the shock front) and $5\times 10^{16}$~cm
wide (crest to trough).  The total intensity of the C~IV and N~V thick
images are $4.8\times 10^6$~photons and $5.6\times 10^6$~photons,
respectively.  The ratio I$_{thin}$/I$_{thick}$ is the factor by which
the observed intensity of a line has to be multiplied to correct for
resonance scattering.  I$_{thin}$ is 10$^7$ for both C~IV and N~V, so
the correction factors for emission from the entire filament are 2.1
and 1.8, respectively.  Note that if the sheet were viewed from above
(or below) rather than edge-on, then I$_{thick}$ would be greater than
I$_{thin}$ for both lines.  This is because the scattered flux
preferentially escapes in the vertical, optically thin direction.

The C~IV and N~V emission from the filament has small scale structure
due to the geometry of the shock front and the stratification of the
post-shock gas.  Therefore, we expect the correction factors to depend
both on position and on the size of the region included.  We first
consider regions of the filament that are $5\times 10^{16}$~cm along
its length and that span its width.  The top panel of
Figure~\ref{fig50unit} shows these regions overlaid on the H$\alpha$
model image.  The middle panel shows plots of the C~IV intensities of
each region, and the bottom panel shows plots of the N~V intensities
of each region.  The intrinsic (thin) and observed (thick)
intensities are shown for both lines.  The position index (along the
$x$-axes) corresponds to the region number as labeled in the H$\alpha$
image.  The intrinsic intensity of both lines is fairly constant for
regions 3 through 10.  In each of these regions, the C~IV intensity is
about $10.6\times 10^5$~photons and the N~V intensity is about
$10.5\times 10^5$~photons.  The observed intensities, which take into
account resonant scattering, are significantly lower.  The C~IV
correction factor is higher than the N~V correction factor for every
region.  The correction factors in regions 3 through 10 range between
1.8 and 3.2 for C~IV and between 1.6 and 2.5 for N~V\@.  For both the
lines, resonance scattering has its maximum effect (i.e., the correction
factor is highest) in region 6, where the filament is the narrowest.

We now consider narrow regions lying parallel to the shock.  In the top
panel of Figure~\ref{figparl} the N~V thick image is shown overlaid
with these regions.  Each region is 100 pixels ($10^{17}$~cm) along the
length of the filament and 6 pixels ($6\times 10^{15}$~cm) wide.  The
regions are numbered, and these numbers are used along the $x$-axis in
the plots.  The middle panel shows plots of the C~IV intensities of
each region and the bottom panel shows plots of the N~V intensities.
The intrinsic (thin) and observed (thick) intensities are shown in each
case.  Of the five selected regions, four are along bright shock
tangencies;  the exception is region 2, where the emission is more diffuse.
This region has the lowest correction factors for both C~IV and N~V\@.
The ratio of I$_{thin}$ to I$_{thick}$ for each line is 1.4 in region 2.
Region 4 lies along the narrowest and brightest part of the filament.
As expected, the correction factors are highest for this region.  The 
correction factor is 3.6 for C~IV and 2.9 for N~V\@.  The N~V emission
from regions 3 and 5 are very similar - $3.0\times 10^5$ and
$3.1\times 10^5$ observed photons, respectively and correction factors
of 2.0 and 2.1.  The C~IV emission from these regions are quite 
different.  There are $1.3\times 10^5$ observed photons from region 3
and $2.5\times 10^5$ observed photons from region 5; the correction
factors are 2.8 and 2.0 for regions 3 and 5, respectively. 

The intensities and correction factors presented so far are for regions
of different sizes with no regard to spatial variations within a
region.  We now consider the spatial structure of the emission across
the filament.  In the top panel of Figure~\ref{figslits}, the H$\alpha$
model is shown, overlaid with three boxes representing spectrograph
slits.  Each box is 3 pixels ($3\times10^{15}$~cm) wide and 45 pixels
long.  The left panels are plots of C~IV intensities and the right
panels are plots of N~V intensities.  Thin and thick intensities have
been plotted as functions of position along the slit.  The bright
tangencies seen in the H$\alpha$ image have corresponding peaks in the
intrinsic (thin) C~IV and N~V intensities.  The leading and trailing
edges of the filament structure are clearly seen for all the slits.
Additionally, for slit $b$, there is a peak at $x\sim 23$.  As expected
from the shock model used (\S2; see also Figure~\ref{figmodim}), the
emission peaks are more pronounced and narrower in C~IV than in N~V\@.
The effect of resonance scattering is to level these peaks and produce
more uniform emission along the slit.  This is evident from the plots
in Figure~\ref{figslits}.  For both lines and all three positions, the
variation in the thick intensity across the slit is less than a factor
of 2 (excluding the ends where the intensity is dropping off to zero).
In contrast, the maximum thin intensities along the slit are factors of
between 3 and 7 higher than the minimum (again excluding the ends).

The correction factors for C~IV and N~V depend upon location within the
slit.  They also depend upon the number of spatial points that are
binned together.  Summing up all the emission in the slit, the
correction factors are as follows:  (a) 1.9 for C~IV; 1.7 for N~V, (b)
2.3 for C~IV; 1.9 for N~V, and (c) 2.4 for C~IV; 2.0 for N~V\@.  The
correction factors do not differ by much among the three slit
positions.  Also, as in the regions considered earlier, the C~IV
correction factor is higher than the N~V correction factor in each
case.  In Figure~\ref{figrats}, the correction factors are plotted as
functions of position along the slit for each of the three slit
positions.  (Although the information here is available in the plots of
Figure~\ref{figslits}, we present it separately for clarity.) The
correction factors for C~IV and N~V are most different from each other
at the bright edges of the filament.  When the edge is more or less
straight, the correction factors are the highest.  For instance, at
$x\sim 30$ in slit $a$ and at $x\sim 15$ in slit $c$, the C~IV
correction factor is $\sim 6$, and the N~V correction factor is $\sim
3$.  The correction factors are lower, and the contrast between C~IV
and N~V is smaller on curved edges ($x\sim 7$ in $a$, $x\sim 12$ in
$b$, and $x\sim 16$ in $c$).


\section{Discussion}

The C~IV and N~V line fluxes are important diagnostics of shock
properties.  In a non-radiative shock, the post-shock gas is in the
process of reaching the highest ionization state it can, and so the
C~IV to N~V flux ratio is very sensitive to both the shock velocity and
the column swept up by the shock.  For instance, in the shock model
that we have used for this paper ($v_s$ = 180~km~s$^{-1}$; n$_0$ =
2~cm$^{-3}$), the C~IV to N~V flux is 2:1 when the swept up column is
$\sim8\times 10^{15}$~cm$^{-2}$ and is 1:1 when the swept up column is
$\sim1.6\times 10^{16}$~cm$^{-2}$.  (The C~IV and N~V fluxes are also
linearly dependent on the carbon and nitrogen abundances,
respectively.) The results of the radiative transfer simulation
(presented in \S3 above) have shown that resonance scattering decreases
the C~IV and N~V fluxes significantly.  The correction factors for
these lines, which we have defined as the ratio of intrinsic to
observed flux, depend upon the location and the size of the region
being observed.  These correction factors vary between about 1.5 and 6
for the cases we have considered.

We can apply our current results to the HUT spectrum of the filament
presented by \cite{lon92} (this is the same filament on which we have
based our model).  The spectrum was obtained through an aperture
9\farcs4~$\times$~116\arcsec, which corresponds to a linear size about
that of the model filament.  Therefore the correction factors for
emission from the entire filament, 2.1 for C~IV and 1.8 for N~V (\S3),
are applicable to the respective line strengths measured in the HUT
spectrum.  Since the detailed morphology of the filament was unknown,
\cite{lon92} underestimated the C~IV and N~V optical depth and did not
correct for resonance line scattering.  It is not our purpose here to
re-analyze the HUT data.  It is, in any case, not possible to do so based on
just two correction factors since there are several resonance lines in
the spectrum.  Furthermore most of the results in \cite{lon92} were
based on flux ratios rather than fluxes, and would not change
drastically.  Our aim is to underscore the point that high spatial
resolution data have made it worthwhile, and even necessary, to
calculate the effect of resonance line scattering accurately.

Even though we have considered radiative transfer models for only one
set of parameters and a single shock geometry, we can infer some
general results for non-radiative shocks.  For instance, when regions
large enough to average over the shock structure are considered, the
C~IV correction factor is larger than the N~V correction factor, but it
is less than twice as large.  Also, the correction factors for both
lines are highest at places where the filament is narrowest.  The
contrast between C~IV and N~V correction factors is a maximum at the
filament edges.  We expect these results to hold for sheet-like shock
geometries for a range of shock conditions where the extent of the C~IV
and N~V zones are comparable to those in the models presented here.  In
order to interpret spatially resolved spectra where different regions
of a single filament are sampled, a model using just one set of input
parameters is insufficient.  Even for a fixed shock geometry (say,
using H$\alpha$ imaging as in the case of the Cygnus Loop filament),
the influence of resonance scattering on line fluxes would depend on
the post-shock structure, and a grid of models for a range of shock
parameters would need to be calculated.

We have been able to specify the geometry of the non-radiative filament
in a straightforward way, since the shock front is smooth.  In
contrast, most of the bright optical filaments in SNRs are associated
with radiative shocks (e.g.~\cite{ray81}, \cite{bla92}) and have
complex morphologies.  These shocks are subjected to thermal
instabilities and the material in the post-shock flow is often knotty
and fragmented.  The Monte Carlo radiative transfer models can be
applied to these shocks only if the density structure of the emitting
gas is prescribed.  The physically correct way to do this would be to
calculate hydrodynamical simulations of the shock interaction.  A
realistic simulation would need to be done in at least two dimensions
to capture the instabilities, and it would have to include cooling,
which plays an essential role in the shock dynamics.


\acknowledgements
We thank the referee for a careful reading of the paper and
constructive comments.  We also thank John Raymond, with whom we had
helpful discussions.  This work has been supported by STScI grant
GO-07289.01-96A to the Johns Hopkins University and by NASA grant
NAG5-6039.


\clearpage

\figcaption{Intensity images for the non-radiative shock model.  (a)
optically thin H$\alpha$ emission, (b) optically thin C~IV emission,
(c) optically thick C~IV emission, (d) optically thin N~V emission and
(e) optically thick N~V emission.  Each image corresponds to a spatial
extent $7\times 10^{17}$~cm by $1\times 10^{17}$~cm.  The shock moves
upward in these figures.  The H$\alpha$ image has been linearly scaled
while each pair of ``thin'' and ``thick'' images (i.e.~for C~IV and
N~V) have been logarithmically scaled between the same minimum and
maximum values for ease of comparison.
            \label{figmodim}}

\figcaption{\textit{Top panel:} the H$\alpha$ model image overlaid with a grid.
Sections between successive vertical (dashed) lines have been summed to
produce the plots in the figure.  Each section is 50 pixels wide,
corresponding to $5\times 10^{16}$~cm.  
\textit{Middle panel:} the C~IV intrinsic (thin) and observed (thick) photon
intensities.  Each point corresponds to one section of the filament.
\textit{Bottom panel:} The N~V thin and thick photon intensities.
The position index, along the $x$-axis in each plot, is the 
section number labeled on the H$\alpha$ model image.
            \label{fig50unit}}

\figcaption{\textit{Top panel:} the N~V thick model image with several
regions overlaid.  Each region is 100 pixels long and 6 pixels wide
($10^{17}$~cm by $6\times 10^{15}$~cm).
\textit{Middle panel:} the C~IV intrinsic (thin) and observed (thick) photon
intensities for each region.
\textit{Bottom panel:} The N~V thin and thick photon intensities.
The region numbers in each plot are shown along the $x$-axis and
correspond to the numbers marked on the N~V image in the top panel.
            \label{figparl}}

\figcaption{At the top, the H$\alpha$ model image is shown overlaid
with three boxes representing spectrograph slits.  Each box is 3 pixels
wide and 45 pixels long.  The boxes are labeled \textit{a, b} and
\textit{c}.  In the plots, the C~IV and N~V intensities are plotted
as a function of position along each of these boxes.  The plots along
the left column show the C~IV thin and thick intensities; the plots
along the right column show the N~V thin and thick intensities.
The origin of the $x-axis$ corresponds to the top of the slit in each
case so the shock direction is right to left in these plots.  The
$y-axis$ scale is the same for all plots for ease of comparison.
            \label{figslits}}

\figcaption{Plots of the correction factors as functions of position
along the slit for the three slit positions shown in
Figure~\protect\ref{figslits}.
            \label{figrats}}

%
%
%
%

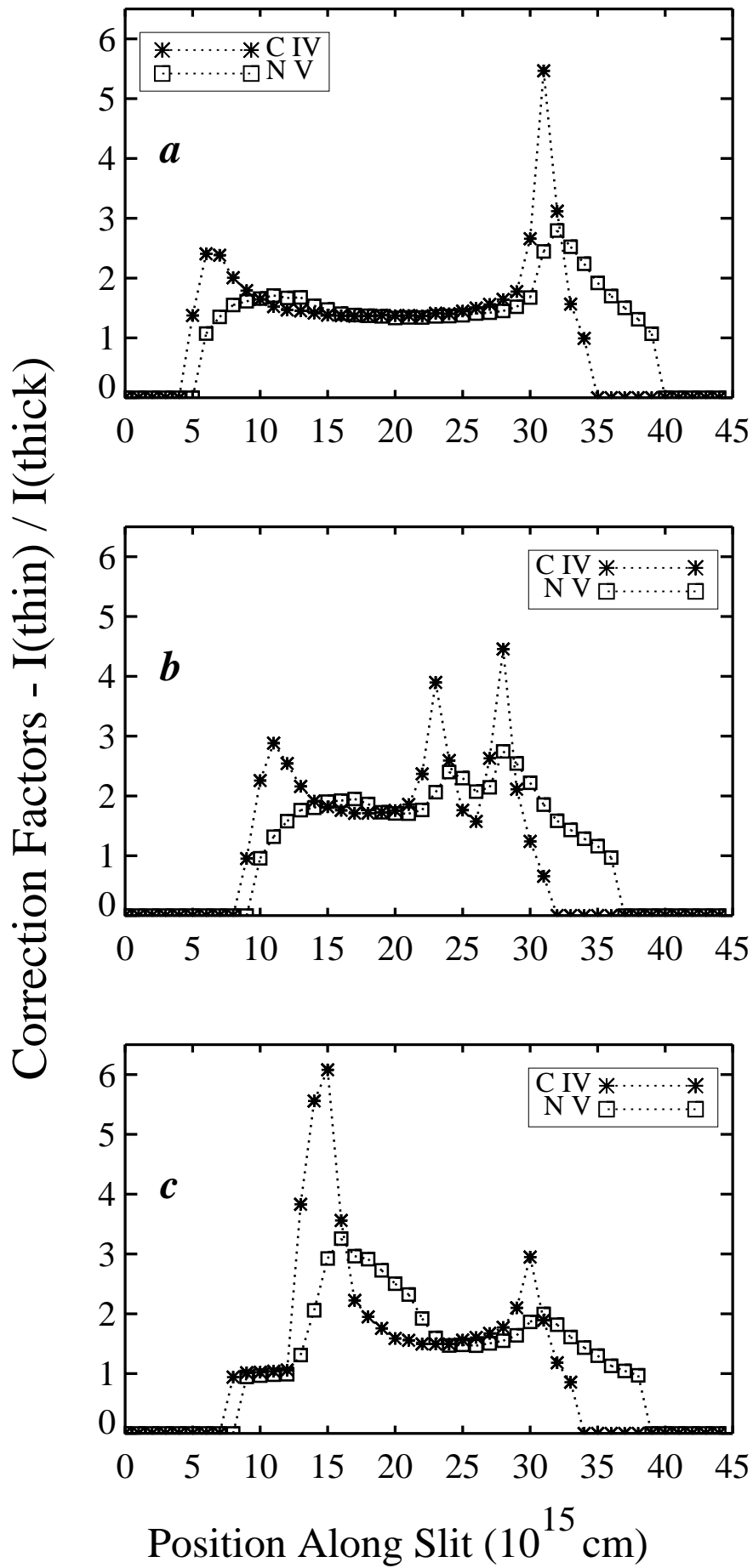
\end{document}